\documentclass[conference]{IEEEtran}

\usepackage{amsmath, amssymb, bm, cite, epsfig, psfrag}
\usepackage{epstopdf}
\usepackage{graphicx}
\usepackage{algorithm,algpseudocode}
\usepackage{array}

\usepackage{bbm}
\usepackage{multirow}
\usepackage[usenames,dvipsnames]{xcolor}
\renewcommand{\arraystretch}{1.5}
\usepackage{subfigure}
\usepackage{etoolbox}
\usepackage{pbox}

\usepackage{colortbl}
\usepackage{fancyhdr}
\usepackage[margin=0.54in]{geometry}
\usepackage[hyphens]{url}
\usepackage[hidelinks]{hyperref}
\hypersetup{breaklinks=true}
\usepackage{cite}
\usepackage[vertfit,hyphenbreaks]{breakurl}
\usepackage[width=0.48\textwidth,font=small]{caption}
\usepackage{capt-of}
\newtoggle{conference}
\togglefalse{conference} 
\graphicspath{{figures/}}

\def\beq{\begin{equation}}
\def\eeq{\end{equation}}
\def\beqa{\begin{eqnarray}}
\def\eeqa{\end{eqnarray}}
\def\beqan{\begin{eqnarray*}}
\def\eeqan{\end{eqnarray*}}

\def\PL{\mathrm{PL}}
\def\dB{\mathrm{dB}}

\def\tm1{t\! - \! 1}
\def\tp1{t\! + \! 1}

\def\PL{\textrm{PL}}
\def\dB{\textrm{dB}}
\def\FSPL{\textrm{FSPL}}

\def\CIF{\textrm{CIF}}

\def\ABG{\textrm{ABG}}

\def\BPL{\textrm{BPL}}

\def\1m{\textrm{1 m}}

\def\PL{\mathrm{PL}}
\def\dB{\mathrm{dB}}

\usepackage{lipsum}
\usepackage{fancyhdr}
\usepackage{datetime}

\usepackage{tikz}
\usetikzlibrary{calc}

\begin{document}

\newcommand\blfootnote[1]{%
  \begingroup
  \renewcommand\thefootnote{}\footnote{#1}%
  \addtocounter{footnote}{-1}%
  \endgroup
}

\pagestyle{empty}


\title{5G 3GPP-like Channel Models for Outdoor Urban Microcellular and Macrocellular Environments}

\author{
	\IEEEauthorblockN{Katsuyuki Haneda\textsuperscript{a}, Lei Tian\textsuperscript{b}, Yi Zheng\textsuperscript{c}, Henrik Asplund\textsuperscript{d}, Jian Li\textsuperscript{e}, Yi Wang\textsuperscript{e}, David Steer\textsuperscript{e}, Clara Li\textsuperscript{f}, \\
		Tommaso Balercia\textsuperscript{f}, Sunguk Lee\textsuperscript{g}, YoungSuk Kim\textsuperscript{g}, Amitava Ghosh\textsuperscript{h}, Timothy Thomas\textsuperscript{h} Takehiro Nakamura\textsuperscript{i},\\
		Yuichi Kakishima\textsuperscript{i}, Tetsuro Imai\textsuperscript{i}, Haralabos Papadopoulas\textsuperscript{i}, Theodore S. Rappaport\textsuperscript{j}, George R. MacCartney Jr.\textsuperscript{j},\\
		 Mathew K. Samimi\textsuperscript{j}, Shu Sun\textsuperscript{j}, Ozge Koymen\textsuperscript{k}, Sooyoung Hur\textsuperscript{l}, Jeongho Park\textsuperscript{l}, Charlie Zhang\textsuperscript{l}, Evangelos Mellios\textsuperscript{m}, \\
		 Andreas F. Molisch\textsuperscript{n}, Saeed S. Ghassamzadeh\textsuperscript{o}, and Arun Ghosh\textsuperscript{o} \\
	\text{\textsuperscript{a}Aalto University}, \textsuperscript{b}BUPT, \textsuperscript{c}CMCC, \textsuperscript{d}Ericsson, \textsuperscript{e}Huawei,  \textsuperscript{f}Intel, \textsuperscript{g}KT Corporation, \textsuperscript{h}Nokia, \textsuperscript{i}NTT DOCOMO,} 
	\text{\textsuperscript{j}NYU WIRELESS, \textsuperscript{k}Qualcomm, \textsuperscript{l}Samsung, \textsuperscript{m}University of Bristol, \textsuperscript{n}University of Southern California, \textsuperscript{o}AT\&T}
}

\maketitle
\begin{tikzpicture} [remember picture, overlay]
\node at ($(current page.north) + (0,-0.25in)$) {K. Haneda \emph{et al.}, ``5G 3GPP-like Channel Models for Outdoor Urban Microcellular and Macrocellular Environments,''};
\node at ($(current page.north) + (0,-0.4in)$) {to be published in \textit{2016 IEEE 83rd Vehicular Technology Conference (VTC 2016-Spring)}, May, 2016.};
\end{tikzpicture}
\begin{abstract}
For the development of new 5G systems to operate in bands up to 100 GHz, there is a need for accurate radio propagation models at these bands that currently are not addressed by existing channel models developed for bands below 6 GHz. This document presents a preliminary overview of 5G channel models for bands up to 100 GHz. These have been derived based on extensive measurement and ray tracing results across a multitude of frequencies from 6 GHz to 100 GHz, and this document describes an initial 3D channel model which includes: 1) typical deployment scenarios for urban microcells (UMi) and urban macrocells (UMa), and 2) a baseline model for incorporating path loss, shadow fading, line of sight probability, penetration and blockage models for the typical scenarios. Various processing methodologies such as clustering and antenna decoupling algorithms are also presented.  
\end{abstract}

\section{Introduction}
Next generation 5G cellular systems will encompass frequencies from around 500 MHz up to 100 GHz. For the development of the new 5G systems to operate in bands above 6 GHz, there is a need for accurate radio propagation models for these higher frequencies that have yet to be addressed. Previous generations of channel models were designed and evaluated for operation at frequencies only as high as 6 GHz. 

One important example is the recently developed 3D-urban micro (UMi) and 3D-urban macro (UMa) channel models for LTE~\cite{3GPP36873}. This paper is a summary of key results provided in a much more detailed white paper by the authors, that can be found at the link in~\cite{5GSIG}. The 3GPP 3D channel model provides additional flexibility for the elevation dimension, thereby allowing modeling for two dimensional antenna systems, such as those that are expected in next generation system deployments. Future system design will require new channel models that will be validated for operation at higher frequencies (e.g., up to 100 GHz) and that will allow accurate performance evaluation of possible future technical specifications for these bands over a representative set of possible environments and scenarios of interest.  These new models should be consistent with the models below 6 GHz. In some cases, the requirements may call for deviations from the modelling parameters or methodology of the existing 3GPP models, but these deviations should be kept to a bare minimum and only introduced when necessary for supporting the 5G simulation use cases.

There are many existing and ongoing campaign efforts worldwide targeting 5G channel measurements and modeling. They include METIS2020~\cite{METIS2015}, COST2100/COST~\cite{COST2100}, IC1004~\cite{COSTic1004}, ETSI mmWave~\cite{ETSI2015}, NIST 5G mmWave Channel Model Alliance~\cite{NIST}, MiWEBA~\cite{MiWEBA2014}, mmMagic~\cite{mmMagic}, and NYU WIRELESS~\cite{Rap13a,Rap15a,Rap15b,Mac15a}. METIS2020, for instance, has focused on 5G technologies and has contributed extensive studies in terms of channel modelling over a wide range of frequency bands (up to 86 GHz), very large bandwidths (hundreds of MHz), and three dimensional polarization modelling, spherical wave modelling, and high spatial resolution. The METIS channel models consist of a map-based model, stochastic model, and a hybrid model which can meet requirements of flexibility and scalability. The COST2100 channel model is a geometry-based stochastic channel model (GSCM) that can reproduce the stochastic properties of multiple-input/multiple output (MIMO) channels over time, frequency, and space. On the other hand, the NIST 5G mmWave Channel Model Alliance is newly established to provide guidelines for measurement calibration and methodology, modeling methodology, as well as parameterization in various environments and a database for channel measurement campaigns. NYU WIRELESS has conducted and published extensive urban propagation measurements at 28, 38, 60, and 73 GHz for both outdoor and indoor channels, and has created large-scale and small-scale channel models, including the concepts of \emph{time cluster spatial lobes} (TSCL) to model multiple multipath time clusters that are seen to arrive in particular directions~\cite{Rap15a,Rap13a,Samimi15a,Samimi15b,Samimi15c}

In this document, we present a brief overview of the outdoor channel properties for bands up to 100 GHz based on extensive measurement and ray tracing results across a multitude of bands. In addition we present a preliminary set of channel parameters suitable for 5G simulations that are capable of capturing the main properties and trends.

\section{Requirements for New Channel Model}
The requirements of the new channel model that will support 5G operation across frequency bands up to 100 GHz should preferably be based on the existing 3GPP 3D channel model~\cite{3GPP36873} but with extensions to cater for additional 5G modeling requirements and scenarios, for example: a) antenna arrays, especially at higher-frequency millimeter-wave bands, will very likely be 2D and dual-polarized both at the access point (AP) and the user equipment (UE) and will hence need properly-modeled azimuth and elevation angles of departure and arrival of multipath components; b) individual antenna elements will have antenna radiation patterns in azimuth and elevation and may require separate modeling for directional performance gains. Furthermore, polarization properties of the multipath components need to be accurately accounted for in the model.

Also, the new channel model must accommodate a wide frequency range up to 100 GHz. The joint propagation characteristics over different frequency bands will need to be evaluated for multi-band operation, e.g., low-band and high-band carrier aggregation configurations.

Furthermore, the new channel model must support large channel bandwidths (up to 2 GHz), where: a) the individual channel bandwidths may be in the range of 100 MHz to 2 GHz and may support a) carrier aggregation, and b) the operating channels may be spread across an assigned range of several GHz.

The new channel model must also support a range of large antenna arrays, in particular: a) some large antenna arrays will have very high directivity with angular resolution of the channel down to around 1.0$^\circ$, b) 5G will consist of different array types, e.g., linear, planar, cylindrical and spherical arrays, with arbitrary polarization, c) the array manifold vector can change significantly when the bandwidth is large relative to the carrier frequency. As such, the wideband array manifold assumption is not valid and new modeling techniques may be required. It may be preferable, for example, to model arrival/departure angles with delays across the array and follow a spherical wave front assumption instead of the usual plane wave assumption.

Additionally, the new channel model must accommodate mobility, in particular: a) the channel model structure should be suitable for mobility up to 350 km/hr, b) the channel model structure should be suitable for small-scale mobility and rotation of both ends of the link in order to support scenarios such as device to device (D2D) or vehicle to vehicle (V2V).

Moreover, the new channel model must ensure spatial/temporal/frequency consistency, in particular: a) the model should provide spatial/temporal/frequency consistencies which may be characterized, for example, via spatial consistence, inter-site correlation, and correlation among frequency bands, b) the model should also ensure that the channel states, such as line-of-sight (LOS)/non-LOS (NLOS) for outdoor/indoor locations, the second order statistics of the channel, and the channel realizations, change smoothly as a function of time, antenna position, and/or frequency in all propagation scenarios, c) the spatial/temporal/frequency consistencies should be supported for simulations where the channel consistency impacts the results (e.g. massive MIMO, mobility and beam tracking, etc.). Such support could possibly be optional for simpler studies.

When building on to existing 3GPP models, the new channel model must be of practical computational complexity, in particular: a) the model should be suitable for implementation in single-link simulation tools and in multi-cell, multi-link radio network simulation tools. Computational complexity and memory requirements should not be excessive. The 3GPP 3D channel model~\cite{3GPP36873} is seen, for instance, as a sufficiently accurate model for its purposes, with an acceptable level of complexity. Accuracy may be provided by including additional modeling details with reasonable complexity to support the greater channel bandwidths, and spatial and temporal resolutions and spatial/temporal/frequency consistency, required for millimeter-wave modeling. b) the introduction of a new modeling methodology (e.g. map based model) may significantly complicate the channel generation mechanism and thus substantially increase the implementation complexity of the system level simulator. Furthermore, if one applies a completely different modeling methodology for frequencies above 6 GHz, it would be difficult to have meaningful comparative system evaluations for bands up to 100 GHz.

\section{Typical UMi and UMa Outdoor Scenarios}

\subsection{UMi Channel Characteristics (TX Heights $<$ 25 m)}
Work by the authors show that LOS path loss in the bands above 6 GHz appear to follow Friis' free space path loss model quite well. Just as in lower bands, a higher path loss slope (or path loss exponent when using a 1 m close in reference distance) is observed in NLOS conditions. The shadow fading in the measurements appears to be similar to lower frequency bands, while ray-tracing results show a much higher shadow fading ($>$ 10 dB) than measurements, due to the larger dynamic range allowed and much greater loss in some ray tracing experiments. In NLOS conditions at frequencies below 6.0 GHz, the RMS delay spread is typically modelled at around 50-500 ns, the RMS azimuth angle spread of departure (from the AP) at around 10$^\circ$ to 30$^\circ$, and the RMS azimuth angle spread of arrival (at the UE) at around 50$^\circ$ to 80$^\circ$~\cite{3GPP36873}. There are measurements of the delay spread above 6 GHz which indicate somewhat smaller ranges as the frequency increases, and some measurements show the millimeter wave omnidirectional channel to be highly directional in nature.

\subsection{UMa Channel Characteristics (TX Heights $\geq$ 25 m)}
Similar to the UMi scenario, the LOS path loss behaves quite similar to free space path loss, as expected. For the NLOS path loss, the trends over frequency appear somewhat inconclusive across a wide range of frequencies. The rate at which the loss increases with frequency does not appear to be linear, as the rate is higher in the lower part of the spectrum. This could possibly be due to diffraction, which is frequency dependent, being a more dominating propagation mechanism at the lower frequencies. At higher frequencies reflections and scattering may be more predominant~\cite{Rap15a}. 

Alternatively, the trends could be biased by the lower dynamic range in the measurements at the higher frequencies. More measurements are needed to understand the UMa channel. From preliminary ray-tracing studies, the channel spreads in delay and angle appear to be weakly dependent on the frequency and are generally 2-3 times smaller than in~\cite{3GPP36873}. The cross-polar scattering in the ray-tracing results tends to increase (lower XPR) with increasing frequency due to diffuse scattering. 

\section{Outdoor-To-Indoor Penetration Loss}
In both the UMa and the UMi scenarios a significant portion of UEs or devices are expected to be indoors. These indoor UEs increase the strain on the link budget since additional losses are associated with the penetration into buildings. The characteristics of the building penetration loss and in particular its variation over the higher frequency range is therefore of high interest and a number of recent measurement campaigns have been targeting the material losses and building penetration losses at higher frequencies~\cite{Zhao13a,Rodriguez14a,Larsson14a}. The current understanding, based on these measurements is briefly summarized as follows.

\begin{figure}
	\centering
	\includegraphics[width = 0.5\textwidth]{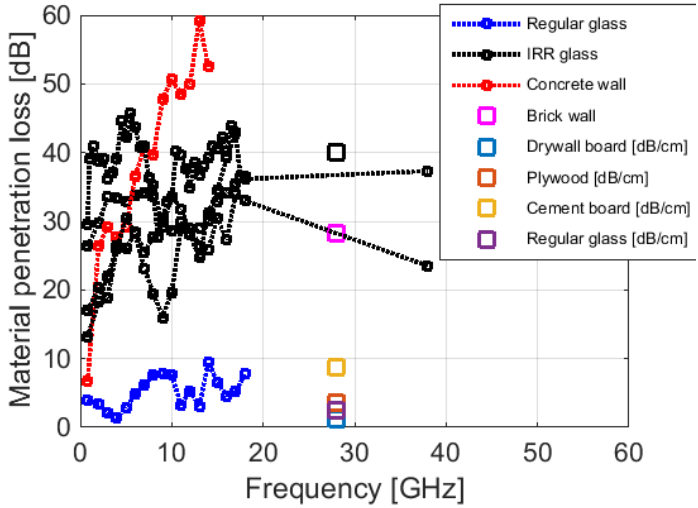}
	\caption{Measured material penetration losses. Sources:~\cite{Zhao13a,Rodriguez14a}, and measurements by Samsung and Nokia.}
	\label{fig:Penetration1}
\end{figure}

Different materials commonly used in building construction have very diverse penetration loss characteristics. Common glass tends to be relatively transparent with a rather weak increase of loss with higher frequency due to conductivity losses. “Energy-efficient” glass commonly used in modern buildings or when renovating older buildings is typically metal-coated for better thermal insulation. This coating introduces additional losses that can be as high as 40 dB even at lower frequencies. Materials such as concrete or brick have losses that increase rapidly with frequency. Fig.~\ref{fig:Penetration1} summarizes some recent measurements of material losses. The loss trends with frequency are linear to a first order of approximation. Variations around the linear trend can be understood from multiple reflections within the material or between different layers which cause constructive or destructive interference depending on the frequency and incidence angle.

\begin{figure}
	\centering
	\includegraphics[width = 0.5\textwidth]{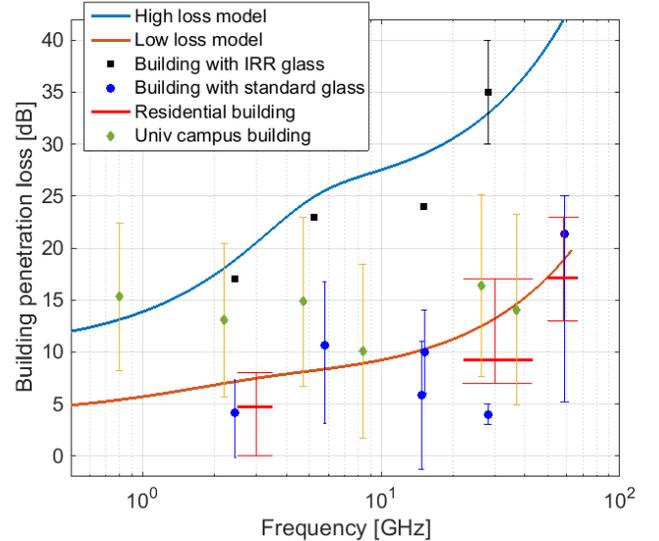}
	\caption{Effective building penetration loss measurements. The bars indicate variability for a given building. Sources:~\cite{Larsson14a} and measurements by Qualcomm, NTT DOCOMO, and Ericsson. The solid curves represent two variants of the model described in~\cite{Semann14a}, which is one out of several penetration loss models. A parabolic curve may also fit the data.}
	\label{fig:Penetration2}
\end{figure}

Typical building facades are composed of several materials, e.g. glass, concrete, metal, brick, wood, etc. Propagation of radio waves into or out of a building will in most cases be a combination of transmission paths through different materials, i.e. through windows and through the facade between windows. The exception could be when very narrow beams are used which only illuminates a single material or when the indoor node is very close to the external wall. Thus, the effective penetration loss can behave a bit differently than the single material loss. A number of recent measurements of the effective penetration loss are summarized in Fig.~\ref{fig:Penetration2}. As indicated by the bars available for some of the measurements, there can be quite some variation even in a single building. For comparison, two models that attempt to capture the loss characteristics of buildings consisting of multiple materials are shown. The loss characteristics of each specific material follows the results shown in Fig.~\ref{fig:Penetration2} quite well which indicates that the results in the material loss measurements and the effective penetration loss measurements are actually fairly consistent even though the loss values behave differently. A parabolic model for building penetration loss (BPL) that fits Fig.~\ref{fig:Penetration2} is:
\begin{equation}\label{BPL}
\BPL [\dB] = 10\log_{10}(A+B\cdot f^2) \text{,}
\end{equation}
where $f$ is frequency in GHz, $A$ = 5, and $B$ = 0.03 for low loss buildings and $A$ = 10 and $B$ = 5 for high loss buildings. 

The majority of the results presented so far have been waves with perpendicular incidence to the external wall. As the incidence angles become more grazing the losses have been observed to increase by up to 15-20 dB. Propagation deeper into the building will also be associated with an additional loss due to internal walls, furniture etc. This additional loss appears to be rather weakly frequency dependent but rather strongly dependent on the interior composition of the building. Observed losses over the 2-60 GHz range from 0.2-2 dB/m

\begin{figure}
	\centering
	\includegraphics[width = 0.5\textwidth]{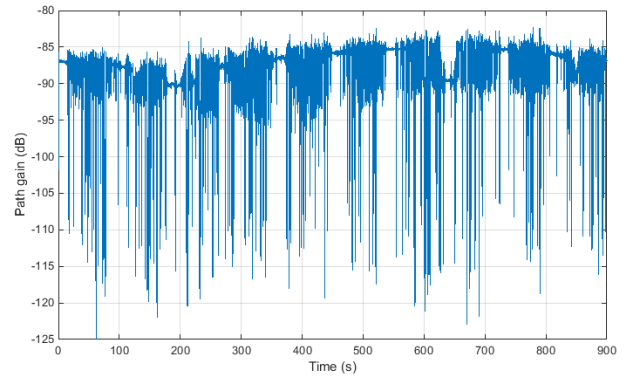}
	\caption{Example of dynamic blockage from a measurement snapshot at 28 GHz.}
	\label{fig:Blockage}
\end{figure}

\section{Blockage Considerations}
\begin{figure}
	\centering
	\includegraphics[width = 0.5\textwidth]{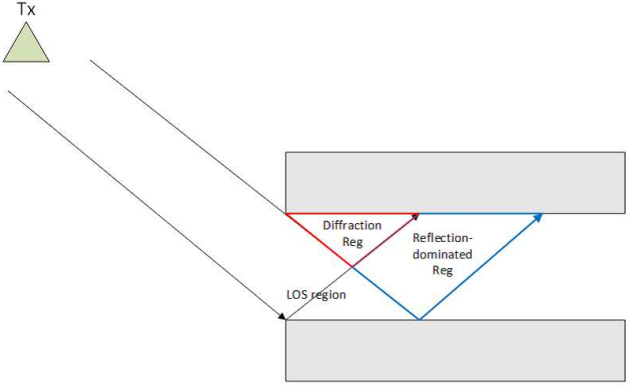}
	\caption{Example of diffraction-dominated and reflection-dominated regions (idealized scenario).}
	\label{fig:Diffraction}
\end{figure}

As the radio frequency increases, its propagation behaves more like optical propagation and may become blocked by intervening objects. Typically, two categories of blockage are considered: dynamic blockage and geometry-induced blockage. Dynamic blockage is caused by the moving objects (i.e., cars, people) in the communication environment. The effect is transient additional loss on the paths that intercept the moving object. Fig.~\ref{fig:Blockage} shows such an example from 28 GHz measurement done by Fraunhofer HHI in Berlin. In these experiments, time continuous measurements were made with the transmitter and receiver on each side of the road that had on-off traffic controlled by a traffic light. Note that the time periods when the traffic light is red is clearly seen in the figure as periods with little variation as the vehicles are static at that time. When the traffic light is green, the blocking vehicles move through the transmission path at a rapid pace as is seen in the figure. The variations seen when the light is red are explained by vehicles turning the corner to pass between the transmitter and receiver. Geometry-induced blockage, on the other hand, is a static property of the environment. It is caused by objects in the map environment that block the signal paths. The propagation channels in geometry-induced blockage locations are dominated by diffraction and sometimes by diffuse scattering. The effect is an exceptional additional loss beyond the normal path loss and shadow fading. Fig.~\ref{fig:Diffraction} illustrates examples of diffraction-dominated and reflection-dominated regions in an idealized scenario. As compared to shadow fading caused by reflections, diffraction-dominated shadow fading could have different statistics (e.g., different mean, variance and coherence distance).

\section{Path Loss, Shadow Fading, LOS and Blockage Modeling}
The LoS state is determined by a map-based approach, i.e., by considering the transmitter (AP) and receiver (UE) positions and whether any buildings or walls are blocking the direct path between the AP and the UE. The impact of objects not represented in the map such as trees, cars, furniture, etc. is modelled separately using shadowing/blocking terms. An attractive feature of this LoS definition is that it is frequency independent, as only buildings and walls are considered in the definition. The first LOS probability model considered, the $d_1$/$d_2$ model, is the current 3GPP/ITU model~\cite{3GPP36873,ITU-M.2135-1}:
\begin{equation}\label{eq1}
p(d) = \min\left(\frac{d_1}{d},1\right)(1-e^{-d/d_2})+e^{-d/d_2}
\end{equation}
where $d$ is the 2D distance in meters and $d_1$ and $d_2$ can both be optimized to fit a set of data (or scenario parameters). The next LOS probability model considered, the NYU (squared) model, is a LOS probability model developed by NYU in~\cite{Samimi15a}:
\begin{equation}\label{eq2}
p(d) = \left(\min\left(\frac{d_1}{d},1\right)(1-e^{-d/d_2})+e^{-d/d_2}\right)^2,
\end{equation}
where again $d_1$ and $d_2$ can be optimized to fit a given set of data (or scenario parameters).

\begin{figure}
	\centering
	\includegraphics[width = 0.5\textwidth]{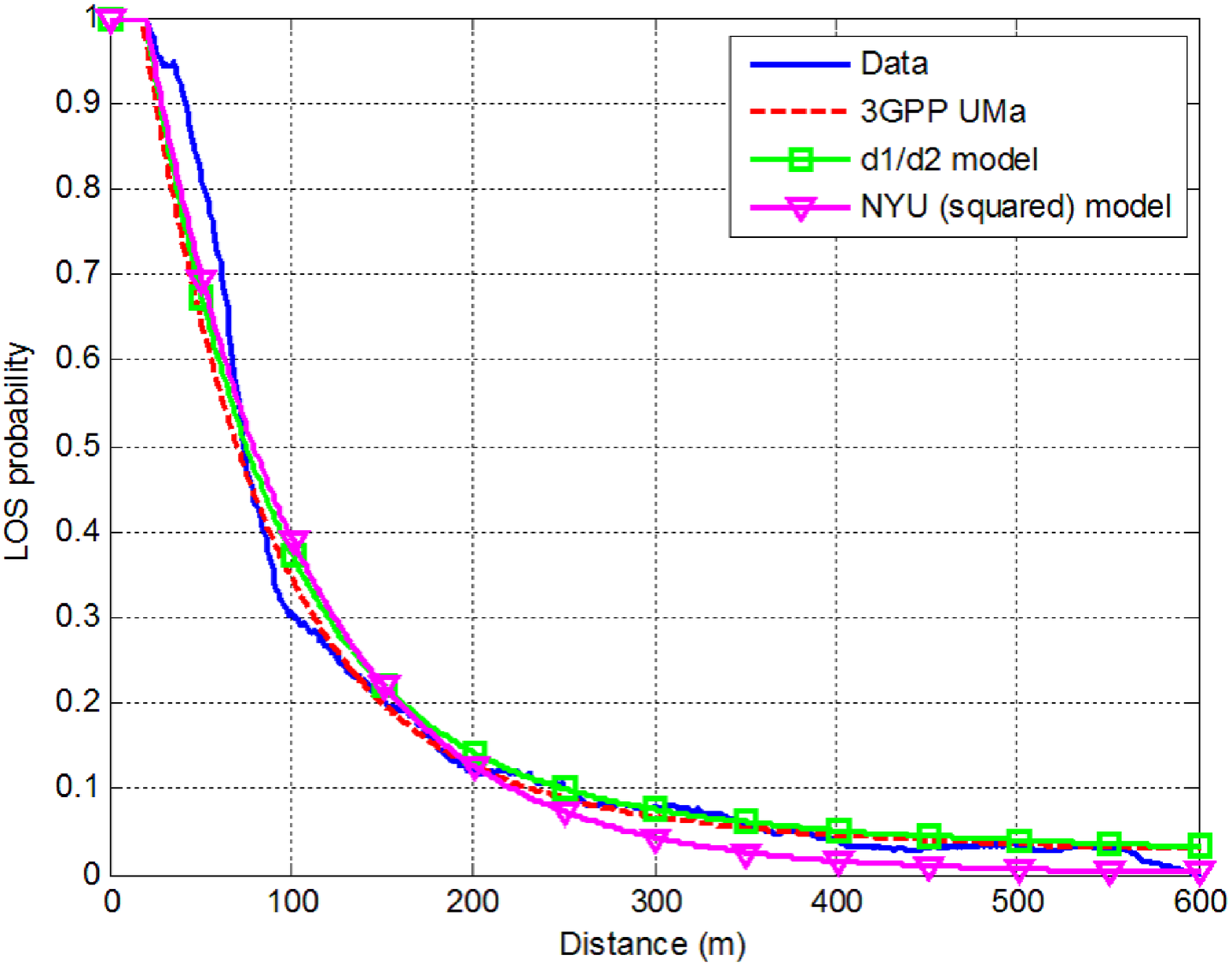}
	\caption{UMa LOS probability for the three models considered.}
	\label{fig:UMaLOS}
\end{figure}

An investigation into the LOS probability for the UMa environment was conducted using all of the UMa measured and ray tracing data. In addition to comparing the two models considered above with optimized $d_1$ and $d_2$ values, the data was also compared to the current 3GPP UMa LOS probability model~\eqref{eq1} for a UE height of 1.5 m with $d_1$ = 18 and $d_2$ = 63. A summary of the results is given in Table~\ref{tbl:UMaLOS} and the three models are compared to the data in Fig.~\ref{fig:UMaLOS}. In terms of mean squared error (MSE) between the LOS probability from the data and the models, the NYU (squared) model had the lowest MSE, but the difference was small. Given that the current 3GPP UMa model was a reasonable match to the data and included support for 3D placement of UEs, it is recommended that the current 3GPP LOS probability model for UMa be used for frequencies above 6.0 GHz. The 3GPP UMa model specifically is~\cite{3GPP36873}:
\begin{equation}\label{eq3}
p(d) = \left(\min\left(\frac{18}{d},1\right)(1-e^{-d/63})+e^{-d/63}\right)\Big(1+C(d,h_{UT})\Big)
\end{equation}
where $h_{UT}$ is the height of the UE in m and:
\begin{equation}\label{eq4}
C(d,h_{UT}) = \begin{cases}
0, & h_{UT}<13\;\text{m}\\
\left(\frac{h_{UT}-13}{10}\right)^{1.5}g(d), & 13\leq h_{UT} \leq 23\; \text{m}
\end{cases}
\end{equation}
\begin{equation}\label{eq5}
g(d) = \begin{cases}
(1.25e^{-6})d^2\exp(-d/150), & d>18\; \text{m}\\
0, & \text{otherwise}
\end{cases}
\end{equation}
Note that for indoor users $d$ is replaced by the 2D distance to the outer wall.

\begin{figure}
	\centering
	\includegraphics[width = 0.5\textwidth]{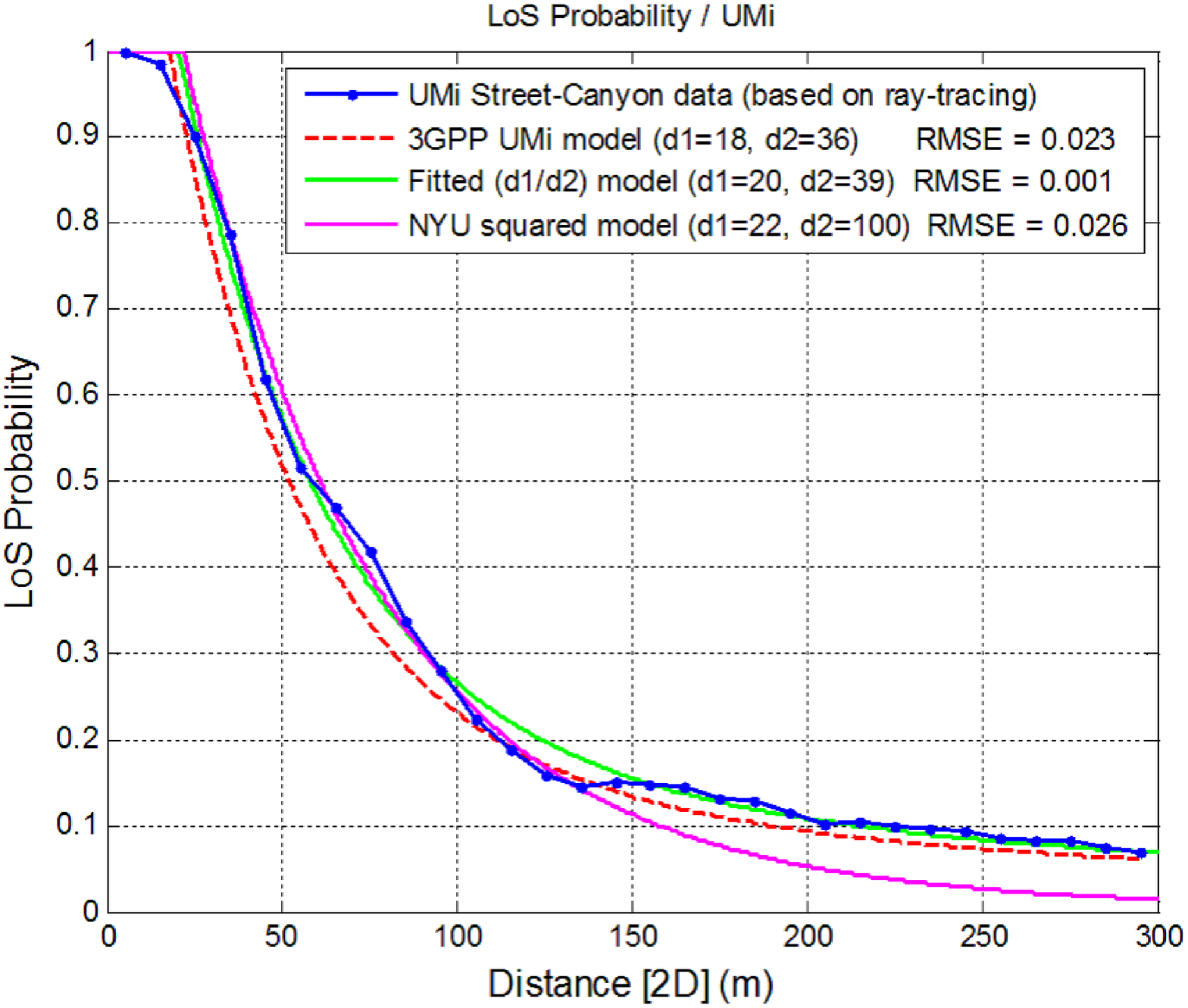}
	\caption{UMi LOS probability for the three models considered.}
	\label{fig:UMiLOS}
\end{figure}
\begin{table}
\caption{Comparison of the LOS probability models for the UMa environment}\label{tbl:UMaLOS}
\centering
\renewcommand{\arraystretch}{1.6}
	\begin{center}
	\scalebox{1}{
	\fontsize{8}{8}\selectfont
	\begin{tabular}{|c||c|c|c|}
		\hline
		& $d_1$ (m) & $d_2$ (m) & MSE \\ \hline \hline
		3GPP UMa & 18 & 63 & 0.020 \\ \hline
		$d_1$/$d_2$ model & 20 & 66 & 0.017 \\ \hline
		NYU (squared) & 20 & 160 & 0.015 \\ \hline
	\end{tabular}}
	\end{center}
\end{table}
\begin{table}
\caption{Comparison of the LOS probability models for the UMi environment}\label{tbl:UMiLOS}
\centering
\renewcommand{\arraystretch}{1.6}
\begin{center}
	\scalebox{1}{
		\fontsize{8}{8}\selectfont
		\begin{tabular}{|c||c|c|c|}
			\hline
			& $d_1$ (m) & $d_2$ (m) & MSE \\ \hline \hline
			3GPP UMa & 18 & 36 & 0.023 \\ \hline
			$d_1$/$d_2$ model & 20 & 39 & 0.001 \\ \hline
			NYU (squared) & 22 & 100 & 0.026 \\ \hline
		\end{tabular}}
	\end{center}
\end{table}
	\begin{table}
\caption{Comparison of the LOS probability models for the UMi environment. S.C. stands for Street Canyon and O.S. stands for Open Square.}\label{tbl:PL}
\centering
\renewcommand{\arraystretch}{1.6}
\begin{center}
	\scalebox{0.82}{
	\fontsize{8}{8}\selectfont
	\begin{tabular}{|c|c|c|}
		\hline
		Scenario & CI Model Parameters & ABG Model Parameters \\ \hline \hline
		UMa-LOS & n = 2.0, SF = 4.1 dB & N/A \\ \hline
		UMa-NLOS & n = 3.0, SF = 6.8 dB & $\alpha = 3.4$, $\beta = 19.2$ $\gamma = 2.3$, SF = 6.5 dB \\ \hline
		UMi-S.C.-LOS & n = 1.98, SF = 3.1 dB & N/A \\ \hline
		UMi-S.C.-NLOS & n = 3.19, SF = 8.2 dB & $\alpha = 3.48$, $\beta = 21.02$ $\gamma = 2.34$, SF = 7.8 dB \\ \hline
		UMi-O.S.-LOS & n = 1.85, SF = 4.2 dB & N/A \\ \hline
		UMi-O.S.-NLOS & n = 2.89, SF = 7.1 dB & $\alpha = 4.14$, $\beta = 3.66$ $\gamma = 2.43$, SF = 7.0 dB \\ \hline
	\end{tabular}}
	\end{center}
\end{table}

For the UMi scenario, it was found that the 3GPP LOS probability formula~\cite{3GPP36873} is sufficient for frequencies above 6 GHz. The fitted $d_1$ = $d_2$ model in~\eqref{eq1} provides a better fitted model, however, the errors between the data and the 3GPP LoS probability model over all distances are small. That formula is the same as in~\eqref{eq1} with $d_1$ = 18 m and $d_2$ = 36 m with $d$ being replaced by the 2D distance to the outer wall for indoor users. Note that the 3GPP UMi LOS probability model is not a function of UE height like the UMa LOS probability model (see Table~\ref{tbl:UMiLOS}).

\subsection{Path Loss Models}
Three multi-frequency PL models are considered here; namely the close-in (CI) free space reference distance PL model~\cite{Rap15b,ITU-M.2135-1,Sun15a,Andersen95a}, the close-in free space reference distance model with frequency-dependent path loss exponent (CIF)~\cite{Mac15a}, and the Alpha-Beta-Gamma (ABG) PL model~\cite{Mac15a,Hata80a,Piersanti12a,Mac13a}. These models are now described and applied to various scenarios.

The path loss models currently used in the 3GPP 3D model is of the ABG model form but without a frequency dependent parameter (AB model) and additional dependencies on base station or terminal height, and with a LOS breakpoint. 3GPP is expected to recommend just one path loss model (per scenario and LOS/NLOS) but that the choice is still open for discussion in 3GPP RAN. Table~\ref{tbl:PL} shows the parameters of the CI and ABG path loss models for different environments for omnidirectional antennas (the CIF model is not used for outdoor modeling as the measurements did not yield a large frequency dependence as observed for indoor measurements~\cite{Mac15a}). It may be noted that the models presented here are multi-frequency models, and the parameters are invariant to carrier frequency and can be applied across the 0.5-100 GHz band.

The CI PL model is given as~\cite{Rap15b,Mac15a}
\begin{equation}\label{eq6}
\PL^{CI}(f,d)[\dB]=\FSPL(f,1\;\text{m})+10n\log_{10}\left(\frac{d}{1\;\text{m}}\right)+X_{\sigma}^{CI}
\end{equation}
where $f$ is the frequency in Hz, $n$ is the PLE, $d$ is the distance in meters, $X_{\sigma}^{CI}$ is the shadow fading (SF) with $\sigma$ in dB, and the free space path loss (FSPL) at 1 m, with frequency $f$ is given as:
\begin{equation}\label{eq7}
\FSPL(f,1\;\text{m})=20\log_{10}\left(\frac{4\pi f}{c}\right),
\end{equation}
where $c$ is the speed of light.

The ABG PL model~\cite{Sun16a,Thomas16a,Piersanti12a,Mac15a} is given as:
\begin{equation}\label{eq8}
\begin{split}
\PL^{\ABG}(f,d)[\dB]=10\alpha\log_{10}(d)+\beta\\
+10\gamma\log_{10}(f)+X^{\ABG}_{\sigma}
\end{split}
\end{equation}
where $\alpha$ captures how the PL increase as the transmit-receive distance (in meters) increases, $\beta$ is a floating offset value in dB, $\gamma$ attempts to capture the PL variation over the frequency $f$ in GHz, and $X^{\ABG}_{\sigma}$  is the SF term with standard deviation in dB.

The CIF PL model is an extension of the CI model~\cite{Mac15a}, and uses a frequency-dependent path loss exponent given by:\begin{equation}\label{eq9}
\begin{split}
\PL^{\CIF}(f,d)[\dB]=\FSPL(f,\text{1 m})+\\10n\Bigg(1+b\left(\frac{f-f_0}{f_0}\right)\Bigg)\log_{10}\left(\frac{d}{\text{1 m}}\right)+X^{\CIF}_{\sigma}
\end{split}
\end{equation}
where $n$ denotes the path loss exponent (PLE), and $b$ is an optimization parameter that captures the slope, or linear frequency dependency of the path loss exponent that balances at the centroid of the frequencies being modeled (e.g., path loss increases as $f$ increases when $b$ is positive). The term $f_0$ is a fixed reference frequency, the centroid of all frequencies represented by the path loss model~\cite{Mac15a}, found as the weighed sum of measurements from different frequencies, using the following equation:
\begin{equation}\label{eq10}
f_0 = \frac{\sum_{k=1}^Kf_k N_K}{\sum_{k=1}^K N_K}
\end{equation}
where $K$ is the number of unique frequencies, and $N_k$ is the  number of path loss data points corresponding to the $k^{th}$  frequency $f_k$. The input parameter $f_0$ represents the weighted frequencies of all measurement (or Ray-tracing) data applied to the model. The CIF model reverts to the CI model when $b$ = 0 for multiple frequencies, or when a single frequency $f$ = $f_0$ is modelled. In the CI PL model, only a single parameter, the path loss exponent (PLE), needs to be determined through optimization to minimize the SF standard deviation over the measured PL data set~\cite{Rap15b,Sun15a,Sun16a}. In the CI PL model there is an anchor point that ties path loss to the FSPL at 1 m, which captures frequency-dependency of the path loss, and establishes a uniform standard to which all measurements and model parameters may be referred. In the CIF model there are 2 optimization parameters ($n$ and $b$), and since it is an extension of the CI model, it also uses a 1 m free-space close-in reference distance path loss anchor. In the ABG PL model there are three optimization parameters which need to be optimized to minimize the standard deviation (SF) over the data set~\cite{Mac15a,Sun16a,Thomas16a}. Closed form expressions for optimization of the model parameters for the CI, CIF, and ABG path loss models are given in~\cite{Mac15a}, where it was shown that indoor channels experience an increase in the PLE value as the frequency increases, whereas the PLE is not very frequency dependent in outdoor UMa or UMi scenarios ($b$ very close to 0 for CIF)~\cite{Rap15b,Sun15a,Sun16a,Thomas16a}. The CI, CIF, and ABG models, as well as cross-polarization forms and closed-form expressions for optimization are given for indoor channels in~\cite{Mac15a}. Table~\ref{tbl:PL} shows the model parameters and shadow factor (standard deviation) from pooled data across several frequency bands from measurements and ray tracing by the authors, as detailed in~\cite{5GSIG}.

\subsection{Fast Fading Model}
\subsubsection{UMi}
In the double-directional channel model, the multipath components are described by the delays and the directions of departure and the direction of arrival. Each multipath component is scaled with a complex amplitude gain. Then the double directional channel impulse response is composed of the sum of the generated double-directional multipath components. The double-directional channel model provides a complete omnidirectional statistical spatial channel model (SSCM) for both LOS and NLOS scenarios in the UMi environment. These results are currently analyzed based on the ray-tracing results, which is compared with the measurement campaign done in the same urban area. The final results will be derived from both the measurement and ray-tracing results. For fast-fading modeling, the ray-tracing based method is useful to extend the sparse empirical datasets and to analyze the channel characteristics in both outdoor and indoor environments.

After the clustering, the results from the ray-tracing simulations are analyzed in the spatio-temporal domain, for cluster parameters such as delays, angles at the TX and RX side, and the received powers. Based on the observed clusters in each link, large-scale parameters such as number of clusters and intra-cluster delay spreads and angle spreads are analyzed using the framework in~\cite{ITU-M.2135-1}, and all parameters are extracted by following the methodologies in~\cite{WINNERII}.

\subsubsection{UMa}
\begin{figure}
	\centering
	\includegraphics[width = 0.5\textwidth]{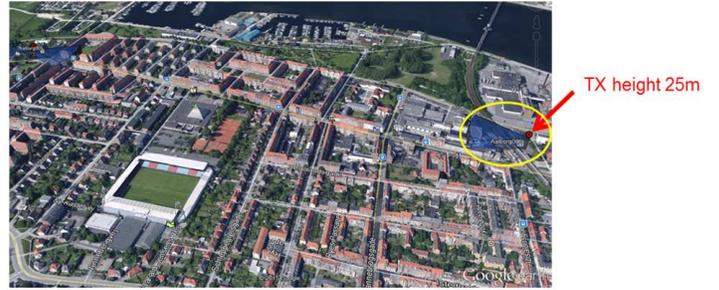}
	\caption{Aalborg, Denmark environment used for ray-tracing study. The AP (TX) location was at the site indicated and UEs were placed outdoors in the streets and open areas.}
	\label{fig:Aalborg}
\end{figure}

Similar to UMi, preliminary UMa large-scale fading parameters in UMa environments were determined using a ray tracing study performed in Aalborg, Denmark as shown in Fig.~\ref{fig:Aalborg}. This environment was chosen as there were real world measurements also made in the same area~\cite{Nguyen16a}. Specifically there was one AP used in the study which had a height of 25 m. The UE height was 1.5 m and isotropic antennas were employed at both the AP and UE. Note that no other objects, such as vehicles, trees, light poles, and signs, were included in this ray tracing study but would be present when measurements were taken. The maximum number of rays in the simulation was 20, no transmissions through buildings were allowed, the maximum number of reflections was four, the maximum number of diffractions was one for frequencies above 10 GHz and was two for frequencies of 10 GHz and below. Six frequencies were considered in this study, i.e., 5.6, 10, 18, 28, 39.3, and 73.5 GHz.

The delay and azimuth angle spreads were found to decrease in frequency. The large-scale parameter that seemed most affected by carrier frequency was the cross-polarization discrimination ratio (XPR), which varied from 13.87 to 7.89 dB when going from 5.6 GHz to 73.5 GHz. The drop in the ray tracing results as frequency increases was primarily attributed to diffuse scattering, as the smaller wavelength of the higher frequency saw an increase in diffuse scattering relative to the lower frequencies, which tends to depolarize the rays. It should be noted that at this point the increasing trend of depolarization at the higher frequencies needs to be verified through measurements. 

Finally, an investigation into the clustering of the rays in this ray-tracing study was performed. To determine clusters, the K-Means algorithm~\cite{Czink06a} was employed with p = 0.98 and s = 0.95 in the shape pruning algorithm. Since this version of the K-Means algorithm has a random starting point (i.e., the first step is a random choosing of the starting centroid positions), the K-Means algorithm was ran 50 times with different random starting points and the cluster set kept at the end was the one which produced the minimum number of clusters. The results showed that the average number of clusters and the average number of rays per cluster were both fairly consistent across the different carrier frequencies.

However, the cluster delay and azimuth angle spreads generally tended to decrease with increasing frequency. In interpreting these results, especially the average number of rays per cluster, it should be noted that the number of modelled rays was limited to 20 in the simulations. More recent 3GPP-like model statistics without such a limitation appear in~\cite{Samimi15c}.

\section{Conclusion}
The basis for this paper is the open literature in combination with recent and ongoing propagation channel measurements performed by a majority of the co-authors of this paper, some of which are as yet unpublished. The preceding tables and figures give an overview of these recent measurement activities in different frequency bands and scenarios. The preliminary findings presented in this paper  and on-going efforts provide promising channel models that can extend today's 3GPP channel models that have been designed for below 6 GHz.

\bibliographystyle{IEEEtran}
\bibliography{5GCMSIG_VTC_V3_4}
\end{document}